\documentclass[%
 aip,
 long, numerical bibliography, (default)
 jmp,%
 amsmath,amssymb,
reprint,%
]{revtex4-1}

\usepackage{graphicx}
\usepackage{dcolumn}
\usepackage{bm}
\begin{document}

\preprint{AIP/123-QED}

\title{Field-effect modulation of conductance in VO$_2$ nanobeam transistors with HfO$_2$ as the gate dielectric}
\author{Shamashis Sengupta}
\email[]{shamashis@tifr.res.in}
\affiliation{Department of Condensed
Matter Physics and Materials Science, Tata Institute of Fundamental
Research, Homi Bhabha Road, Mumbai 400005, India}
\author{Kevin  Wang}
\author{Kai Liu}
\affiliation{Department of Materials Science and Engineering,
University of California, Berkeley, CA 94720, USA}
\author{Ajay K. Bhat}
\author{Sajal Dhara}
\affiliation{Department of Condensed Matter Physics and Materials
Science, Tata Institute of Fundamental Research, Homi Bhabha Road,
Mumbai 400005, India}
\author{Junqiao Wu}
\affiliation{Department of Materials Science and Engineering,
University of California, Berkeley, CA 94720, USA}
\author{Mandar M. Deshmukh}
\affiliation{Department of Condensed Matter Physics and Materials
Science, Tata Institute of Fundamental Research, Homi Bhabha Road,
Mumbai 400005, India}
\date{\today}

\begin{abstract}
We study field-effect transistors realized from VO$_2$ nanobeams
with HfO$_2$ as the gate dielectric. When heated up from low to high
temperatures, VO$_2$ undergoes an insulator-to-metal transition. We
observe a change in conductance ($\sim$ 6 percent) of our devices
induced by gate voltage when the system is in the insulating phase.
The response is reversible and hysteretic, and the area of
hysteresis loop becomes larger as the rate of gate sweep is slowed
down. A phase lag exists between the response of the conductance and
the gate voltage. This indicates the existence of a memory of the
system and we discuss its possible origins.
\end{abstract}

\maketitle

VO$_2$ undergoes an insulator-to-metal transition accompanied by a
change in its crystal structure\cite{eyert,berglund}, the mechanism
of which is still under debate. The transition temperature of a free
crystal is 341 K. Its proximity to room temperature has motivated
attempts at fabricating Mott field-effect transistors (FETs) to
induce the phase transition by applying a gate voltage. Such
experiments have so far been conducted on thin films of VO$_2$
\cite{kim newj, stefanovich, vasilev, ramanathangating}. Other
interesting applications of VO$_2$ include memory
metamaterials\cite{driscollscience} and
memristors\cite{driscollmemristor}. Recently it has been realized
that single-crystalline VO$_2$ nanobeams support single or ordered
metal-insulator domains in the phase transition\cite{junqiao_first,
cao}. This eliminates the random, percolative domain structures
occurring in thin films, and allows intrinsic transition physics to
be probed. In this letter, we report on electrostatic gating
measurements on single crystalline VO$_2$ beams \cite{junqiao_first,
cobden} using HfO$_2$ as the gate dielectric. The devices have a
hysteretic response and appear to possess a memory persisting over a
large timescale (a few minutes). The field effect studies have been
done at different temperatures in the insulating and metallic phases
of the system.

\begin{figure}
\includegraphics[width=80mm, bb=0 0 220.5 187.5]{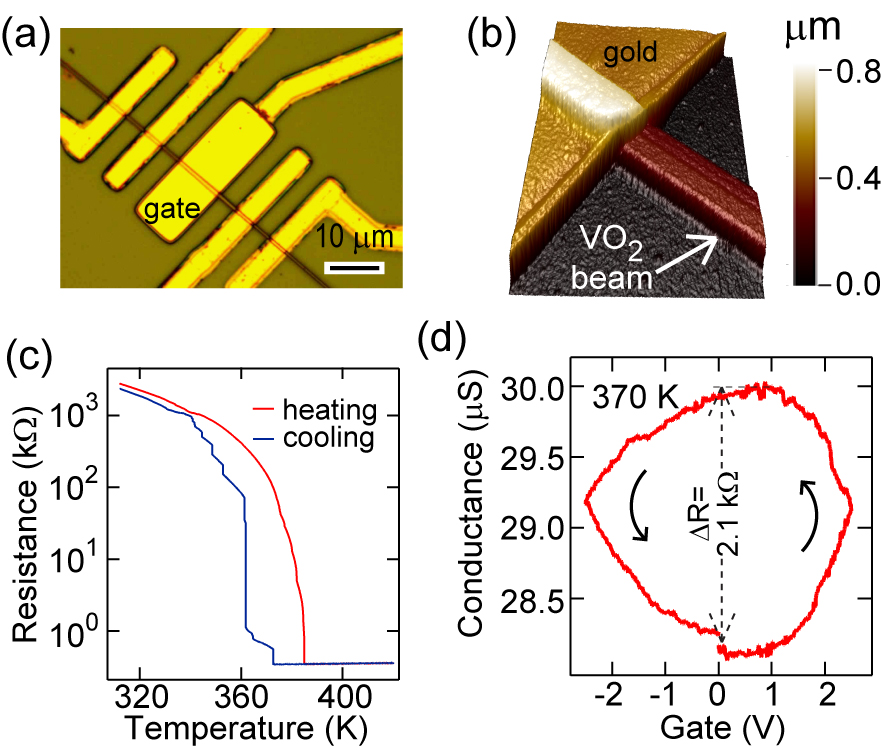}
\caption{\label{fig:figure2} (Color online) (a) Optical microscope
image of a VO$_2$ device. (b) Atomic force microscope image of a
VO$_2$ device. (c) Resistance (in logscale) as a function of
temperature for Device 1. The steps in the cooling curve indicate
metal-to-insulator transition of individual domains. (d) Conductance
of VO$_2$ as a function of gate voltage (data from Device 2). The
resistance $R$, at 0 V to start with,  is 35.4 k$\Omega$.}
\end{figure}

The VO$_2$ beams were grown using the vapor transport
technique\cite{junqiao_first, supp}. Electrodes were designed by
electron beam lithography followed by etching in Ar plasma (for
removal of organic residue) and sputtering of Cr/Au to make Ohmic
contacts. Figs. 1(a) and 1(b) show the optical microscope and atomic
force microscope (AFM) images of VO$_2$ devices. The local gate
electrode in the middle (Fig. 1(a)) is fabricated by first depositing
a 20 nm layer of HfO$_2$ by atomic layer deposition and then
sputtering Cr/Au on top. The typical width of the beams is 0.3-1
$\mu$m, and the thickness is 300-600 nm. Fig. 1(c) shows the
resistance of a VO$_2$ beam as a function of temperature (data from
Device 1). Stress builds up in the system as it is heated, and the
system breaks up into alternating insulator and metal domains
\cite{cao}. The metal domains first appear close to 341 K and on
further heating, grow in size and number. The system becomes
completely metallic at a much higher temperature. The temperature at
which the system turns metallic varies from one device to another
(380-400 K), and is dictated by the stress induced due to adhesion
to the substrate. (The nanobeams are embedded in a 1.1 $\mu$m thick layer of SiO$_2$ grown on Si wafers.)

Two and four probe gating experiments were done inside an evacuated
variable temperature probe station. Both two and four probe
resistances of the same devices were measured (at various
temperatures in both the insulating and metallic phases) and found
to be similar. This indicates that the contact resistance is
negligible compared to the intrinsic resistance of VO$_2$. We have
also confirmed that there is no leakage through the gate\cite{supp}.
Fig. 1(d) shows the effect of gate voltage on the two-probe
conductance of a VO$_2$ device (Device 2) at 370 K. The dc gate
voltage is swept slowly in a cycle (of duration 20 mins) with limiting values of -2.5 and
2.5 V. (The source-drain current used was set at an ac frequency and
monitored with a lock-in amplifier.) Arrows indicate the direction of gate voltage sweep. The
response of the conductance is hysteretic. Gate sweeps at different
rates were conducted on the devices, with the following observation:
the hysteresis loop area and maximum change in conductance become
larger on making the rate of gate sweep slower. This is surprising and has been confirmed on several devices.

\begin{figure}
\includegraphics[width=80mm, bb=0 0 952.0 1097.0]{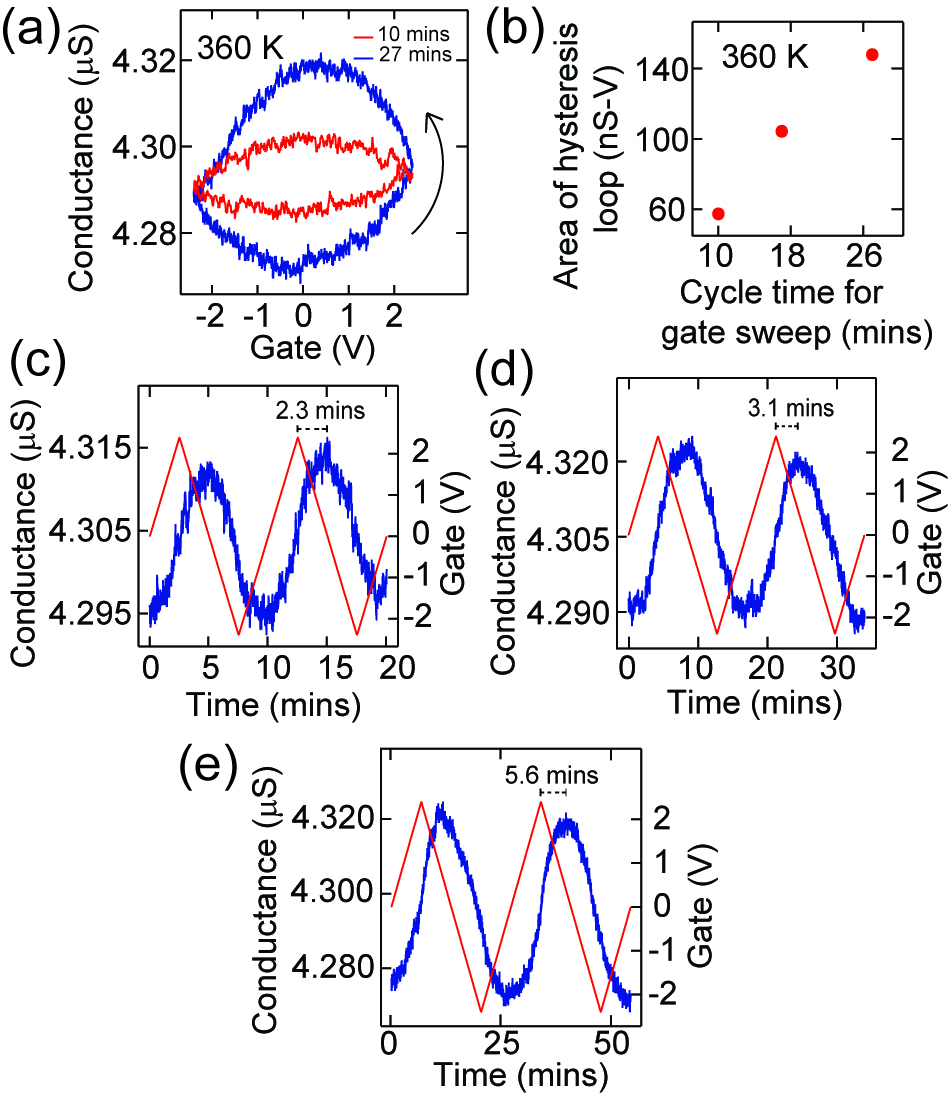}
\caption{\label{fig:figure2} (Color online) (a) Conductance as a
function of gate voltage (Device 3) for two different cycle times:
10 mins (red) and 27 mins (blue). (Note: The former (red curve) is
offset by -0.009 $\mu$S) (b) Area of `conductance vs. gate voltage'
hysteresis loop at different cycle times. (c), (d), (e) Gate
voltage (red) and conductance (blue) plotted against time for cycle
times 10 mins, 17 mins and 27 mins respectively.}
\end{figure}

Fig. 2(a) shows two probe conductance ($G$) as a function of gate
voltage ($V_g$) at 360 K for Device 3 at different gate voltage
sweep-rates. The cycle which is swept slowly over 27 mins (blue
curve) has a much larger hysteresis than the one which is swept
faster (red curve) in 10 mins. The area of the loop is computed as
$\sum G \Delta V_g$ where the summation extends over one cycle of
gate voltage. In Fig. 2(b), it is shown how the area of the loop
increases with an increase in the cycle time (i.e., slowing down of
the gate voltage sweep-rate). Another intriguing aspect is
prominently seen in Figs. 1(d) and 2(a). As we increase $V_g$ up from 0
V to higher positive values (see Fig. 1(d)), $G$ increases. At the
extreme value of 2.5 V, $V_g$ is reversed backwards. However, $G$
does not start reducing immediately. It goes on increasing for a
while and starts to reduce only after a time lag. (Denoting time as
$t$, we can say that $\frac{dG}{dt}$ does not change sign
simultaneously with $\frac{dV_g}{dt}$.) This implies that the system
wants to persist in the state of `increasing conductance' even
though the gate voltage has reversed. This is a manifestation of the
`memory' or `inertia' of the system. This memory effect \cite{diventra} is observed
at the other extreme of gate voltage (-2.5 V) also. The gate voltage
and resulting conductance (data from Device 3) are plotted
simultaneously as a function of time in Figs. 2(c)-(e). (Each plot shows
two consecutive cycles of gate voltage.) In all these curves, it is
seen that the maximum (minimum) of conductance is shifted in time
from the maximum (minimum) of gate voltage. This shift, or `phase
lag' between the input and output signals, is the signature of a
persistent effect. Slower the rate of sweep, larger is the
time-delay. It is 5.6 mins for the slowest scan with a 27 mins cycle
(Fig. 2(e)).

\begin{figure}
\includegraphics[width=80mm, bb=0 0 240.5 214.75]{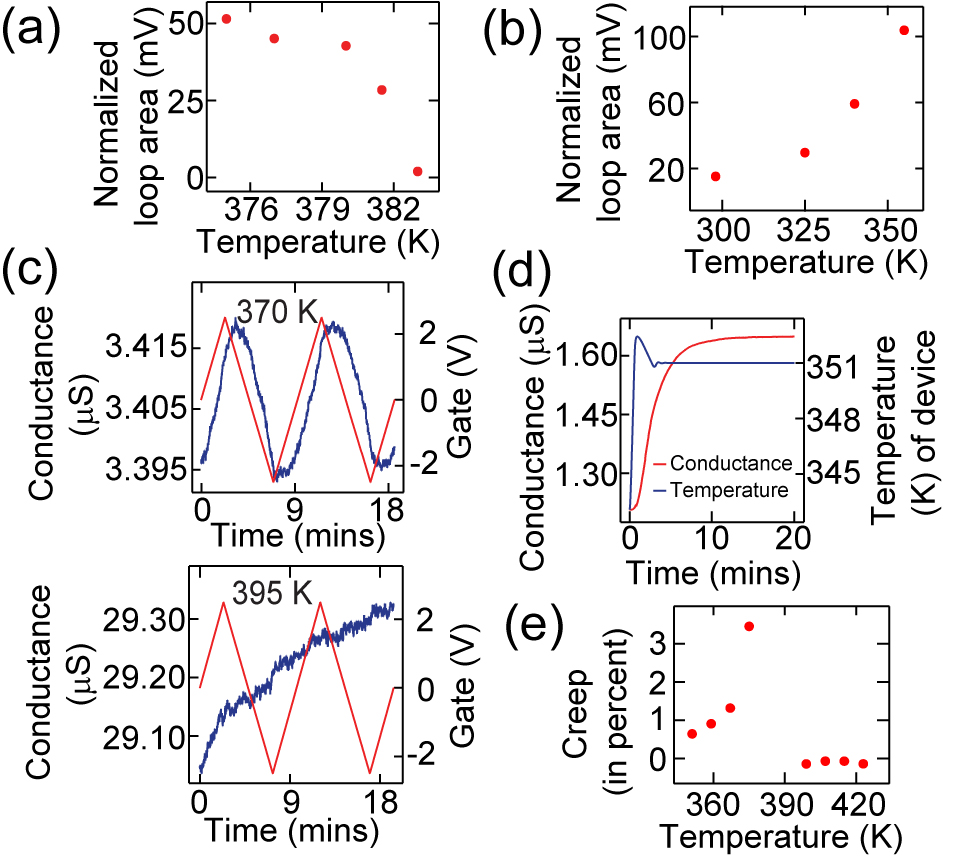}
\caption{\label{fig:figure2} (Color online) (a) Normalized loop area
of `conductance vs. gate voltage' hysteresis close to the
insulator-metal transition around 383 K (data from Device 3). Cycle time of gate voltage sweep is 25
mins. (b) Normalized loop area of `conductance vs. gate voltage' hysteresis at
different temperatures in the insulating state of Device 1. This was
a four probe measurement and time for each gate voltage cycle was 7
mins. (c) Gate voltage (red) and conductance (blue) plotted against
time at two different temperatures in the insulating phase of Device
4. (Cycle time is 9 mins). (d) Temperature of the sample (Device 5)
is ramped up rapidly from 343 K to 351 K. The sample temperature reaches 351
K in 5 mins, but the conductance keeps on increasing slowly over
several minutes after that. (e) Thermal `creep' of Device 5 as a
function of temperature.}
\end{figure}

The hysteresis is observed at temperatures at which the
beam is in the insulating state, or there is a co-existence of metal
and insulator domains\cite{junqiao_first}. No gating is observed in
the full metallic state. We compute the `normalized loop area'
$\frac{\sum G \Delta V_g}{G_{0}}$, where $G_0$ is the conductance at
$V_g$=0. The `normalized loop area' as a function of temperature
(close to the metallic transition) for Device 3 is plotted in Fig.
3(a). The most prominent hysteresis for our devices is usually
obtained in the temperature range 340-370 K, which is the
temperature window in which multiple domains exist along the
beam\cite{junqiao_first, cao}. Also, it is shown in Fig. 3(b) how the
`normalized loop area' varies over a wide range of temperatures
(starting from room temperature) for Device 1.

Fig. 3(c) shows the gate voltage response (as a time chart) for Device
4 at two temperatures. At 370 K, the gate effect ($G$ periodic with
$V_g$) is observed. At 395 K, the VO$_2$ beam is closer to the full
metallic transition and the gate effect has disappeared. However,
there is a gradual variation of the conductance with time. This is
the phenomenon of thermal `creep' that we see in our devices. The
conductance takes a long time to stabilize after the device is
heated to a new temperature. This feature is noticed on all our
devices and is illustrated in Fig. 3(d) (Device 5). The sample is
heated up from 343 K, and it reaches the desired temperature of 351
K within 5 mins. However, even 15 minutes after that, the
conductance of VO$_2$ has not stabilized. It goes on increasing at a
slow rate. (The fractional change over the last 10 mins is 0.64
percent.) We define a quantity called `creep' as the fractional
change in conductance over a period of 10 mins after the sample has
reached a new temperature. The variation with temperature of this quantity is plotted in Fig. 3(e). `Creep'
becomes quite large just before the metallic transition.

The overall change in $G$ is a few percent ($\sim$6 percent in Fig.
1(d) and 1 percent in Fig. 2(e)). Since the entire length of the wire is
not covered by the gate, the fractional change in the gated region
of Device 2 (Fig. 1(d)) turns out to be 14.4 percent\cite{supp}. The
gate voltage primarily affects the carrier density close to the
surface within the surface skin layer, the bulk being
electrostatically screened from the gate. The threshold carrier
concentration\cite{junqiao cao} in VO$_2$ has been estimated to be
$8\times10^{18}$ cm$^{-3}$. Using this value, it is estimated that
the amount of carriers induced by a gate voltage of 2.5 V is 8.3
percent of the intrinsic concentration. This is close (in terms of
order of magnitude) to the fractional change in conductance due to
gating. Hysteretic gating effects are known to arise in
semiconductors due to the presence of surface states at the
dielectric interface. These act as trapping centers for electrons.
It is generally observed that on slowing down the rate of gate
voltage sweep, the system is allowed time to equilibrate and
hysteresis reduces\cite{trap hysteresis}. Hysteresis due to slow
traps (with relaxation time of a few minutes) have also been
reported\cite{slow traps, kingston}. But, in the aforementioned
cases, the observed behavior on varying the sweep rate is the
opposite of what we see in our devices. Hence, trap states do not
seem to offer a possible explanation in our experiments.

Persistent effects have been observed in earlier studies on VO$_2$
(in two terminal memristive devices\cite{driscollmemristor} and
infrared response of gated VO$_2$ films\cite{qazilbash}). In our
experiments, there is no gate leakage\cite{supp} and hence, heating can be ruled out as a
possible cause behind the persistent effect. There is not
much information in literature about mechanical relaxation in
VO$_2$. It is probable that mechanical relaxation time in VO$_2$ is
quite large. When heated to a new temperature, it would take a
considerable period of time for the stress pattern and the relative
domain sizes (and hence, conductance) to settle down. This explains
the thermal `creep'. The VO$_2$ crystal has electric dipoles with
antiferroelectric coupling\cite{junqiao cao}. The coupling strength
will depend upon the spatial separation between the lattice sites,
thus providing a coupling between the dipolar arrangement and the
strain state. Hence, the gate voltage will also affect the strain
state, and relaxation of the dipolar arrangement will have a similar
timescale as the mechanical relaxation. This may explain the slow
processes leading to the time-delay in gate effects (Figs. 2(c)-(e)).

In summary, we have fabricated three terminal field effect devices
from VO$_2$ nanobeams using HfO$_2$ as the dielectric. We observe
gate effects in conductance and the response is hysteretic. The
dependence of electrostatic gating effects on the sweep rate and a
phase lag between the reversal of conductance and gate voltage
indicates that our devices have an intrinsic memory with a large
timescale of a few minutes. This is interesting from the point of
view of probing the physical origin of persistent effect in the
insulating phase of VO$_2$. Also, single crystalline nanobeams with
a smaller thickness may exhibit more pronounced electrostatic gating
effects and can have important implications in the design of Mott
FETs and memory devices.

We thank S. Ramanathan and K. L. Narasimhan for discussions. We
acknowledge the U. S. Department of Energy Early Career Award
DE-0000395 (J. W.) and the Government of India and AOARD-104141 (M. M. D.).

\end{document}